\documentclass[a4paper,12pt,english]{article}
\usepackage[english]{babel}
\usepackage[utf8]{inputenc}
\usepackage{amsmath,setspace,geometry,lineno,amsfonts}
\DeclareMathOperator*{\argmax}{arg\,max}

\usepackage{natbib,authblk,bbm,bm}
\usepackage[boxed]{algorithm}
\usepackage{caption}
\usepackage{algpseudocode}
\usepackage{etoolbox} 
\AtBeginEnvironment{algorithmic}{\scriptsize}
\AtBeginEnvironment{tabular}{\footnotesize}
\AtBeginEnvironment{tabular}{}
\usepackage{graphicx}
\usepackage{multicol}
\usepackage[center]{subfigure}
\usepackage{tikz}
\usetikzlibrary{fit,positioning,arrows,automata,decorations.pathreplacing}

\title{Detecting bearish and bullish markets in financial time series using hierarchical hidden Markov models}

\author{\large Lennart Oelschläger$^{1}\footnote{Corresponding author; email: \texttt{lennart.oelschlaeger@uni-bielefeld.de}; postal address: Universitätsstraße 25, 33615 Bielefeld, Germany}$ ~and Timo Adam$^{1,2}$\\
$^1$Bielefeld University, Germany\\
$^2$University of St Andrews, UK}

\begin{document}

\begin{spacing}{1.25}
\maketitle
\end{spacing}
\vspace{-8.5mm}
\begin{spacing}{1.5}

\begin{abstract}
Financial markets exhibit alternating periods of rising and falling prices. Stock traders seeking to make profitable investment decisions have to account for those trends, where the goal is to accurately predict switches from bullish towards bearish markets and vice versa. Popular tools for modeling financial time series are hidden Markov models, where a latent state process is used to explicitly model switches among different market regimes. In their basic form, however, hidden Markov models are not capable of capturing both short- and long-term trends, which can lead to a misinterpretation of short-term price fluctuations as changes in the long-term trend. In this paper, we demonstrate how hierarchical hidden Markov models can be used to draw a comprehensive picture of financial markets, which can contribute to the development of more sophisticated trading strategies. The feasibility of the suggested approach is illustrated in two real-data applications, where we model data from two major stock indices, the Deutscher Aktienindex and the Standard \& Poor's 500.
\end{abstract}


\section{Introduction}

Earning money with stock trading is simple: one only needs to buy and sell shares at the right moment. In general, stock traders seek to invest at the beginning of upward trends (hereon termed as bullish markets) and repel their shares just in time before the prices fall again (hereon termed as bearish markets). As stock prices depend on a variety of environmental factors \citep{hum09, coh13}, chance certainly plays a fundamental role in hitting those exact moments. However, investigating market behavior can lead to a better understanding of how trends alternate and thereby increases the chance of making profitable investment decisions. By applying the hierarchical extension of hidden Markov models (HMMs), this paper aims at contributing to those investigations. As an illustrating example, we use the proposed methodology to detect bearish and bullish markets in two major stock indices, the Deutscher Aktienindex (DAX) and the Standard \& Poor's 500 (S\&P 500).

Over the last decades, HMMs have emerged as popular tools for modeling financial time series. \cite{ryd98} and \cite{bul06}, for instance, used HMMs to derive stylized facts of stock returns, while \cite{has05} developed an HMM for stock market forecasting. More recently, \cite{lih17} applied HMMs to the S\&P 500, where HMMs were used to identify different levels of market volatility, aiming at providing evidence for the conjecture that returns exhibit negative correlation with volatility. Another application to the S\&P 500 can be found in \cite{ngu18}, where HMMs were used to predict monthly closing prices and to derive an optimal trading strategy, which was shown to outperform the conventional buy-and-hold strategy. All these applications demonstrate that HMMs constitute a versatile class of time series models that naturally accounts for stock markets dynamics.

However, in their basic form HMMs operate on a single time scale, with observations made e.g.\ on a quarterly, monthly, or daily basis. As a consequence, conventional HMMs are not capable of simultaneously capturing short- and long-term trends. In addition, the temporal resolution of the data strongly determines the kind of inference that can be made. This is to be regarded as a major deficit, as short-term price fluctuations can easily be misinterpreted as changes in the long-term trend and hence may draw a distorted picture of market behavior. In this paper, we demonstrate that this issue can to some extent be overcome by adding an hierarchical structure to the basic HMM, which improves the model's ability to distinguish between short- and long-term trends, respectively.

The hierarchical generalization of HMMs originates from supervised machine learning. In \cite{fin98}, hierarchical HMMs (HHMMs) were first applied to solve handwriting recognition tasks, where hierarchical levels appear as single letters, syllables, and words. More recently, \cite{leo17} and \cite{ada19} used HHMMs for animal movement modeling, where the hierarchical structure was exploited to jointly infer animals' behavioral modes at coarse and fine scales, respectively. Here, we demonstrate that the same model structure has also strong potential for modeling financial time series, where it is often of major interest to distinguish short-term (e.g.\ daily) fluctuations from long-term dynamics (e.g.\ bullish and bearish markets lasting several months).

The paper is structured as follows: In Section \ref{sec2}, we introduce the methodology, discuss how HHMMs can be fitted via numerical likelihood maximization, and briefly outline related topics such as state decoding and model checking. In Section \ref{sec3}, we illustrate the feasibility of the suggested approach by modeling time series from the DAX and the S\&P 500, respectively. In Section \ref{sec4}, we conclude with a short discussion and outline possible avenues for future research. Pseudo-code of the proposed methodology is appended, while an implementation in R and C++ is provided in the online appendix.

\section{Methodology}
\label{sec2}

In this section, we introduce the model formulation and dependence structure, state a formula for the likelihood, and discuss its numerical maximization to obtain estimates for the model parameters. In addition, we outline the Viterbi algorithm, which allows to decode the hidden states underlying the observations. Model checking based on pseudo-residuals is discussed at the end of the section. 

\subsection{Model formulation and dependence structure}
HMMs constitute a versatile class of statistical models for time series, see e.g.\ \cite{zuc16} for a comprehensive introduction. They predicate that the behavior of the nature can be divided into a finite number of states, where the state that is active cannot be directly observed. However, at each point in time, a data point is observed, which depends on the current state of the nature and thus yields information on the latter. More formally, this concept can be formulated by introducing two stochastic processes:
\begin{enumerate}
	\item At each time point $t$ of the discrete time space $\{1,\dots,T\}$, an underlying process $(S_t)_t$ selects one state from the state space $\{1,\dots,N\}$. We call $(S_t)_t$ the hidden state process.
	\item Depending on which state is active at $t$, one of $N$ distributions $f^{(1)},\dots,f^{(N)}$ generates the observation $X_t$. The process $(X_t)_t$ is called the observed state-dependent process. 
\end{enumerate}
We make the following assumptions on these processes:
\begin{enumerate}
	\item We assume that $(S_t)_t$ is a time-homogeneous Markov process of first order. The process is therefore identified by its initial distribution $\delta$ and its transition probability matrix (t.p.m.) $\Gamma$.
	\item The process $(X_t)_t$ is said to satisfy the conditional independence assumption, i.e.\ conditionally on the current state $S_t$, the observation $X_t$ is independent of all other states and observations.
\end{enumerate}

From a practical point of view, it is reasonable to identify the initial distribution of $(S_t)_t$ with its stationary distribution $\pi$ (which we assume to exist): 
On the one hand, the hidden state process has been evolving for some time before we start to observe it and hence can be assumed to be stationary. On the other hand, $\pi$ is determined by $\Gamma$ through the equation $\pi\Gamma=\pi$, where setting $\delta=\pi$ reduces the number of parameters that need to be estimated, which is convenient from a computational perspective \citep{zuc16}. 

In case of financial data, the hidden states can be interpreted as different moods of the market. These moods cannot be observed directly. Even though these moods cannot be observed directly, price changes --- which clearly depend on the current mood of the market --- can be observed. Thereby, using an underlying Markov process, we can detect which mood is active at any point in time and how the different moods alternate. Depending on the current mood, a price change is generated by a different distribution. These distributions characterize the moods in terms of expected return and volatility. 

The HMM can be extended by an hierarchical structure, resulting in the HHMM. Throughout this paper, two hierarchies are considered. Assume that we are dealing with two time series observed on two different time scales. For each observation of the time series on the coarser scale, we have several observations of the times series on the finer scale, e.g.\ monthly observations and corresponding daily observations. Following the concept of HMMs, we can model both state-dependent time series jointly. First, we treat the time series on the coarser scale as stemming from an ordinary HMM, which we refer to as the coarse-scale HMM:
\begin{enumerate}
	\item At each time point $t$ of the coarse-scale time space $\{1,\dots,T\}$, an underlying process $(S_t)_t$ selects one state from the coarse-scale state space $\{1,\dots,N\}$. We call $(S_t)_t$ the hidden coarse-scale state process.
	\item Depending on which state is active at $t$, one of $N$ distributions $f^{(1)},\dots,f^{(N)}$ realizes the observation $X_t$. The process $(X_t)_t$ is called the observed coarse-scale state-dependent process. 
\end{enumerate}
The processes $(S_t)_t$ and $(X_t)_t$ have the same properties as before, namely $(S_t)_t$ is a first-order Markov process and $(X_t)_t$ satisfies the conditional independence assumption. 

\begin{figure}[t]
    \centering
        \begin{tikzpicture}
        \tikzstyle{cs}=[circle, minimum size = 15mm, thick, draw = black!80]
        \tikzstyle{csrec}=[rectangle, minimum size = 15mm,  draw=none]
        \tikzstyle{csdots}=[circle, minimum size = 15mm, draw = none]
        \tikzstyle{csinrec}=[circle, minimum size = 15mm, thick, draw = black!80]
        \tikzstyle{csdotsinrec}=[circle, minimum size = 15mm, draw = none]
        \tikzstyle{connect}=[-latex, thick]
          
        \node[csdots] (St-2) [] {$\ldots$};
        \node[cs] (St-1) [right= 20pt of St-2] {$S_{t-1}$};
        \node[cs] (St) [right=85pt of St-1] {$S_{t}$};
        \node[cs] (St+1) [right=85pt of St] {$S_{t+1}$};
        \node[csdots] (St+2) [right=20pt of St+1] {$\ldots$};
        
        \node[csdots] (Xt-2) [above=20pt of St-2] {};
        \node[cs] (Xt-1) [right=of Xt-2, above= 20pt of St-1] {$X_{t-1}$};
        \node[cs] (Xt) [right=of Xt-1, above= 20pt of St] {$X_{t}$};
        \node[cs] (Xt+1) [right=of Xt, above= 20pt of St+1] {$X_{t+1}$};
        \node[csdots] (Xt+2) [right=of Xt+1, above= 20pt of St+2] {};
        
        \node[csdots] (S*t-2t) [below=20pt of St-2] {};
        \node[csdots] (S*t-1t) [right=of S*t-2t, below= 20pt of St-1] {$\ldots$};
        \node[csrec,scale=0.8] (S*tt) [right=of S*t-1t, below=20pt of St] {
        
        \tikz{
        \node[csdotsinrec] (S*tt-2) []  {$\ldots$};
        \node[csinrec] (S*tt-1) [right= 20pt of S*tt-2]  {$S^*_{t,t^*-1}$};
        \node[csinrec] (S*tt) [right= 20pt of S*tt-1]  {$S^*_{t,t^*}$};
        \node[csinrec] (S*tt+1) [right= 20pt of S*tt]  {$S^*_{t,t^*+1}$};
        \node[csdotsinrec] (S*tt+2) [right= 20pt of S*tt+1]  {$\ldots$};
        \node[csinrec] (X*tt-1) [above= 20pt of S*tt-1]  {$X^*_{t,t^*-1}$};
        \node[csinrec] (X*tt) [right= 20pt of X*tt-1, above= 20pt of S*tt]  {$X^*_{t,t^*}$};
        \node[csinrec] (X*tt+1) [right=of X*tt, above= 20pt of S*tt+1]  {$X^*_{t,t^*+1}$};
        
        \path (S*tt-2) edge [connect] (S*tt-1);
        \path (S*tt-1) edge [connect] (S*tt);
        \path (S*tt) edge [connect] (S*tt+1);
        \path (S*tt+1) edge [connect] (S*tt+2);
        \path (S*tt-1) edge [connect] (X*tt-1);
        \path (S*tt) edge [connect] (X*tt);
        \path (S*tt+1) edge [connect] (X*tt+1);
        
        \draw[decorate, decoration={brace,raise=100pt,amplitude=5pt}, thick]  (S*tt-2) -- (S*tt+2);
        }
        
        };
        
        \node[csdots] (S*t+1t) [right=of S*tt, below=of St+1] {$\ldots$}; 
        
        \path (St-2) edge [connect] (St-1);
        \path (St-1) edge [connect] (St);
        \path (St) edge [connect] (St+1);
        \path (St+1) edge [connect] (St+2);
        \path (St-1) edge [connect] (Xt-1);
        \path (St) edge [connect] (Xt);
        \path (St+1) edge [connect] (Xt+1);
        \path (St-1) edge [connect] (S*t-1t);
        \path (St) edge [connect] (S*tt);
        \path (St+1) edge [connect] (S*t+1t);
        \end{tikzpicture}
    \caption{Dependence structure of the HHMM.}
    \label{fig:hhmm}
\end{figure}
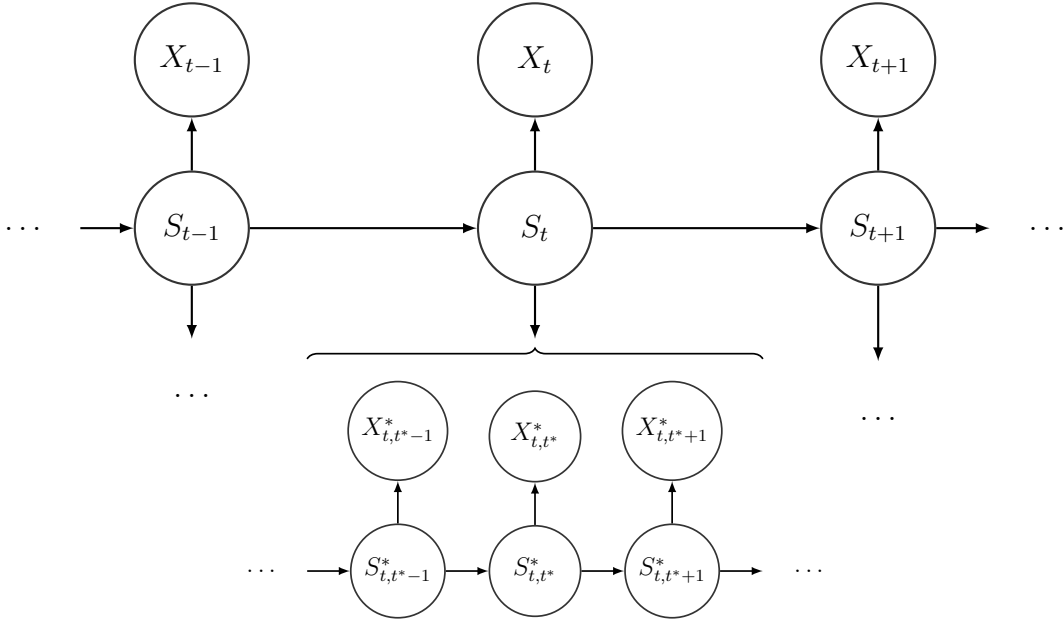

Subsequently, we segment the observations of the fine-scale time series into $T$ distinct chunks, each of which contains all data points that correspond to the $t$-th coarse-scale time point. Assuming that we have $T^*$ fine-scale observations on every coarse-scale time point, we face $T$ chunks comprising of $T^*$ fine-scale observations each. The hierarchical structure now evinces itself as we model each of the chunks by one of $N$ possible fine-scale HMMs. Each of the fine-scale HMMs has its own t.p.m.\ $\Gamma^{*(i)}$, initial distribution $\delta^{*(i)}$, stationary distribution $\pi^{*(i)}$, and state-dependent distributions $f^{*(i,1)},\dots,f^{*(i,N^*)}$. Which fine-scale HMM is selected to explain the $t$-th chunk of fine-scale observations depends on the hidden coarse-scale state $S_t$. The $i$-th fine-scale HMM explaining the $t$-th chunk of fine-scale observations consists of the following two stochastic processes:
\begin{enumerate}
	\item At each time point $t^*$ of the fine-scale time space $\{1,\dots,T^*\}$, the process $(S^*_{t,t^*})_{t^*}$ selects one state from the fine-scale state space $\{1,\dots,N^*\}$. We call $(S^*_{t,t^*})_{t^*}$ the hidden fine-scale state process.
	\item Depending on which state is active at $t^*$, one of $N^*$ distributions $f^{*(i,1)},\dots,f^{*(i,N^*)}$ realizes the observation $X^*_{t,t^*}$. The process $(X^*_{t,t^*})_{t^*}$ is called the observed fine-scale state-dependent process. 
\end{enumerate}
The fine-scale processes $(S^*_{1,t^*})_{t^*},\dots,(S^*_{T,t^*})_{t^*}$ and $(X^*_{1,t^*})_{t^*},\dots,(X^*_{T,t^*})_{t^*}$ satisfy the Markov property and the conditional independence assumption, respectively, as well. Furthermore, it is assumed that the fine-scale HMM explaining $(X^*_{t,t^*})_{t^*}$ only depends on $S_t$. The dependence structure of the HHMM with two hierarchies is visualized in Figure~\ref{fig:hhmm}.

\subsection{Likelihood evaluation and numerical maximization}

Conceptually, an HHMM can be treated as an HMM with two conditionally independent observations, the coarse-scale observation on the one hand and the corresponding chunk of fine-scale observations connected to a fine-scale HMM on the other hand. To derive the likelihood of an HHMM, we start by stating the likelihood formula for the fine-scale HMMs. 

Assume that we want to fit the $i$-th fine-scale HMM, with model parameters $\theta^{*(i)}=(\delta^{*(i)},\Gamma^{*(i)},(f^{*(i,k)})_k)$, to the $t$-th chunk of fine-scale observations, $(X_{t,t^*})_{t^*}$. Consider the so-called fine-scale forward probabilities 
$\alpha^{*(i)}_{k,t^*}=f^{*(i)}(X^*_{t,1},\dots,X^*_{t,t^*}, S^*_{t,t^*}=k)$,
where $t^*=1,\dots,T^*$ and $k=1,\dots,N^*$. Obviously,
\begin{align*}
\mathcal{L}^\text{HMM}(\theta^{*(i)}\mid (X^*_{t,t^*})_{t^*})=\sum_{k=1}^{N^*}\alpha^{*(i)}_{k,T^*}.
\end{align*}
The forward probabilities can be calculated in a recursive way of linear complexity:
\begin{align*}
\alpha^{*(i)}_{k,1} = \delta^{*(i)}_k f^{*(i,k)}(X^*_{t,1}) \quad \text{and} \quad
\alpha^{*(i)}_{k,t^*} = f^{*(i,k)}(X^*_{t,t^*})\sum_{j=1}^{N^*}\gamma^{*(i)}_{jk}\alpha^{*(i)}_{j,t^*-1}, \quad t^*=2,\dots,T^*.
\end{align*}

The transition from the likelihood function of an HMM to the likelihood function of an HHMM is straightforward: Consider the so-called coarse-scale forward probabilities
$\alpha_{i,t}=f(X_1,\dots,X_t,(X^*_{1,t^*})_{t^*},\dots,(X^*_{t,t^*})_{t^*}, S_t=i)$,
where $t=1,\dots,T$ and $i=1,\dots,N$. The likelihood function of the HHMM results from these variables as
\begin{align*}
\mathcal{L}^\text{HHMM}(\theta,(\theta^{*(i)})_i\mid (X_t)_t,((X^*_{t,t^*})_{t^*})_t)=\sum_{i=1}^{N}\alpha_{i,T}.
\end{align*}
The coarse-scale forward probabilities can be calculated in a similar way by applying the recursive scheme
\begin{align*}
\alpha_{i,1} &= \delta_i \mathcal{L}^\text{HMM}(\theta^{*(i)}\mid (X^*_{1,t^*})_{t^*})f^{(i)}(X_1), \\
\alpha_{i,t} &= \mathcal{L}^\text{HMM}(\theta^{*(i)}\mid (X^*_{t,t^*})_{t^*}) f^{(i)}(X_t)\sum_{j=1}^{N}\gamma_{ji}\alpha_{j,t-1}, \quad t=2,\dots,T.
\end{align*}

Maximization of the likelihood function is numerically feasible using the Newton-Raphson method. In practice, we often face the technical issues such as numerical under- or overflow, which can be addressed by maximizing the logarithm of the likelihood and incorporating constants in a conducive way. Instead of computing the forward probabilities directly, we consider the logarithmic transformation 
$\phi^{*(i)}_{k,t^*}=\log[\alpha^{*(i)}_{k,t^*}]$ and $\phi_{i,t}=\log[\alpha_{i,t}]$
thereof (log-forward probabilities). The recursive form described above remains: The fine-scale log-forward probabilities satisfy
\begin{align*}
\phi^{*(i)}_{k,1}&=\log[\delta^{*(i)}_k]+\log[f^{*(i,k)}(X^*_{t,1})], \\
\phi^{*(i)}_{k,t^*}&=\log[f^{*(i,k)}(X^*_{t,t^*})]+\log\left[\sum_{j=1}^{N^*} \gamma^{*(i)}_{jk} \exp[ \phi^{*(i)}_{j,t^*-1} -c_{t^*-1}]\right]+c_{t^*-1},
\end{align*}
where $c_{t^*-1}=\max \{ \phi^{*(i)}_{1,t^*-1},\dots,\phi^{*(i)}_{N^*,t^*-1} \}$ and $t^*=2,\dots,T^*$. The log-likelihood of a fine-scale HMM results from these variables as 
\begin{align*}
\log \mathcal{L}^\text{HMM}(\theta^{*(i)}\mid (X^*_{t,t^*})_{t^*})=\log \left[ \sum_{k=1}^{N^*}\exp[\phi^{*(i)}_{k,T^*}-c_{T^*}]\right]+c_{T^*},
\end{align*}
where $c_{T^*} = \max\{ \phi^{*(i)}_{1,T^*},\dots,\phi^{*(i)}_{N^*,T^*} \}$. See Algorithm \ref{alg:llhmm} in the appendix for pseudo-code of the computation. Similarly, the coarse-scale log-forward probabilities satisfy
\begin{align*}
\phi_{i,1}&=\log[\delta_i]+\log \mathcal{L}^\text{HMM}(\theta^{*(i)}\mid (X^*_{1,t^*})_{t^*})+\log[f^{(i)}(X_{1})], \\
\phi_{i,t}&=\log \mathcal{L}^\text{HMM}(\theta^{*(i)}\mid (X^*_{t,t^*})_{t^*})+\log[f^{(i)}(X_t)]+\log\left[\sum_{j=1}^{N}\gamma_{ji}\exp[\phi_{j,t-1}-c_{t-1}]\right]+c_{t-1},
\end{align*}
where $c_{t-1} = \max\{ \phi_{1,t-1},\dots,\phi_{N,t-1} \}$ and $t=2,\dots,T$. The log-likelihood of the HHMM results from these variables as 
\begin{align*}
\log \mathcal{L}^\text{HHMM}(\theta,(\theta^{*(i)})_i\mid (X_t)_t,((X^*_{t,t^*})_{t^*})_t)=\log\left[\sum_{i=1}^{N}\exp[\phi_{i,T}-c_{T}]\right]+c_{T},
\end{align*}
where $c_{T} = \max\{ \phi_{1,T},\dots,\phi_{N,T} \}$. See Algorithm \ref{alg:llhhmm} in the appendix for a pseudo-code. 

Additionally, we have to consider that certain model parameters must satisfy constraints, namely the transition probabilities and potentially parameters of the state-dependent distributions. Using parameter transformations serves the purpose. To ensure that the entries of the t.p.m.s fulfill non-negativity and the unity condition, we use a bijective transformation from the real numbers to the unity interval. Rather than estimating the probabilities $(\gamma_{ij})_{i,j}$ directly, we estimate unconstrained values $(\eta_{ij})_{i\neq j}$ for the non-diagonal entries of $\Gamma$ and derive the probabilities using the multinomial logit link
\begin{align*}
\gamma_{ij}=\frac{\exp[\eta_{ij}]}{1+\sum_{k\neq i}\exp[\eta_{ik}]},~i\neq j.
\end{align*}
The diagonal entries result via the unity condition $\gamma_{ii}=1-\sum_{j\neq i}\gamma_{ij}$. Noteworthy, not $N^2$ but $N(N-1)$ parameters have to be estimated for an $N\times N$-t.p.m. Furthermore, variances are strictly positive, which can be achieved by applying an exponential transformation to the unconstrained estimator.

A third source of conflicts arises from the fact that the likelihood is maximized with respect to a relatively large number of parameters, which can lead to local maxima apart from the global maximum. Common Newton-Raphson-type optimization routines are unable to distinguish local maxima from the global one. To avoid the trap of ending up at a local maximum, it is recommended to run the maximization routine multiple times from different, possibly randomly selected starting points, and to choose the model that corresponds to the highest likelihood. The number of runs should increase with the number of parameters.

\subsection{State decoding}
\label{sec2:3}

In practice, the primary interest often lies in decoding the hidden states. The term $\argmax_{S_1,\dots,S_T} f(S_1,\dots,S_T\mid X_1,\dots,X_T)$ represents the most-likely underlying state sequence $(S_t)_t$ of an HMM given the data $(X_t)_t$, which is equivalent to the expression $\argmax_{S_1,\dots,S_T} f(S_1,\dots,S_T, X_1,\dots,X_T)$. This in turn can be computed using the Viterbi algorithm, see \cite{zuc16}. The algorithm is based on the variables
\begin{align*}
\xi_{i,t} &= \max_{S_1,\dots,S_{t-1}} f(S_1,\dots,S_{t-1},S_t=i,X_1,\dots,X_t),
\end{align*}
$t=1,\dots,T$ and $i=1,\dots,N$, which can be calculated recursively via
\begin{align*}
\xi_{i,1} = \delta_i f^{(i)}(X_1) \quad \text{and} \quad \xi_{i,t} = \max_j \left(  \xi_{j,t-1}\gamma_{ji}  \right)f^{(i)}(X_t).
\end{align*}
Obtaining the most-likely state sequence $(\hat{S}_t)_t$ is feasible using these variables, starting at the end of the time horizon and going backwards in time:  
\begin{align*}
\hat{S}_T = \argmax_i \xi_{i,T}  \quad \text{and} \quad \hat{S}_t = \argmax_i \xi_{i,t}\gamma_{i\hat{S}_{t+1}},~t=T-1,\dots,1.
\end{align*}
As for the likelihood function, we need to prevent numerical conflicts. Therefore, we again apply a logarithmic transformation, see Algorithm $\ref{alg:viterbi}$ in the appendix, where $\kappa_{i,t}=\log [ \xi_{i,t} ]$. State decoding in HHMMs is straightforward by first decoding the coarse-scale process and using this information to decode the fine-scale process afterwards, see \cite{ada19}.

\subsection{Model checking} 
Analyzing so-called pseudo-residuals enables us to check whether a fitted HMM describes the data sufficiently well. This cannot be done by standard residual analysis since the observations are explained by different distributions, depending on the active state. Therefore, all observations have to be transformed on a common scale in the following way: If $X_t$ has the invertible distribution function $F_{X_t}$, then $Z_t=\Phi^{-1}(F_{X_t} (X_t))$ is standard normally distributed, where $\Phi$ denotes the cumulative distribution function of the standard normal distribution. The observations, $(X_t)_t$, are modeled well if the pseudo-residuals, $(Z_t)_t$, are approximately standard normally distributed. More generally for HHMMs, we first decode the coarse-scale state process using the Viterbi algorithm (see Section \ref{sec2:3}). Subsequently, we assign each coarse-scale observation its associated distribution function under the fitted model and perform the transformation described above. Using the decoded coarse-scale states, we treat the fine-scale observations analogously.

\section{Application to stock market indices}
\label{sec3}
Stock market indices are typically computed as weighted averages over the share prices of several companies in the market, thereby constituting a convenient measure of the overall market behavior. In this section, HHMMs are applied to the DAX and the S\&P 500\footnote{The data were downloaded from www.finance.yahoo.com on July 13, 2020.}, pursuing the goal of detecting long-term trends along with short-term trends.

\subsection{DAX}
\label{sec3-1}
The DAX averages the share prices of the 30 largest publicly traded companies in Germany. Formally, its value $I_t$ at time point $t$ equals
\begin{align*}
	I_t = \frac{\sum_{i} p_{i,t}\cdot q_{i,t^*}\cdot c_{i,t}}{\sum_{i} p_{i,t_0}\cdot q_{i,t_0}}\cdot K_{t^*}\cdot 1000,
\end{align*} 
where $i=1,\dots,30$ denotes the included companies, $t_0$ is the basis date (December 30, 1987), $t^*$ denotes the time point of the last adjustment, $p_{i,t}$ is the share price and $q_{i,t}$ denotes the capital of company $i$ at time point $t$, respectively, and $c_{i,t}$ and $K_{t^*}$ are adjustment factors, see \cite{jan92}. 

Instead of modeling the process $(I_t)_t$ directly, we consider the time series
\begin{align*}
	X_t = \log\left[  \frac{I_t}{I_{t-1}} \right], ~ t\geq 2,
\end{align*}
which we refer to as the logarithm of the daily returns (log-returns). This transformation yields two important benefits from a modeling point of view: First, conditional independence becomes a reasonable assumption, which is required to preserve the first-order Markov property of the hidden state process. Second, financial theory suggests a distribution for log-returns, namely the $t$-distribution, see \cite{pla08}. 

For our application, the fine-scale observation process is identified by the daily log-returns of the years 2000 to 2020. For the coarse-scale observations, we incorporate the average log-returns over 30 trade days\footnote{While this choice is somewhat arbitrary, we found that choosing different fine-scale time horizons between 20 and 40 days had no major impact on the estimation results.}, which appears to be a reasonable time span at which short-term trends can manifest themselves. AIC and BIC favor 3 coarse-scale states and 2 fine-scale states. 

\begin{figure}[t]
	\centering
	\subfigure[Estimated state-dependent distributions of daily log-returns.]{
		\includegraphics[width=0.45\textwidth]{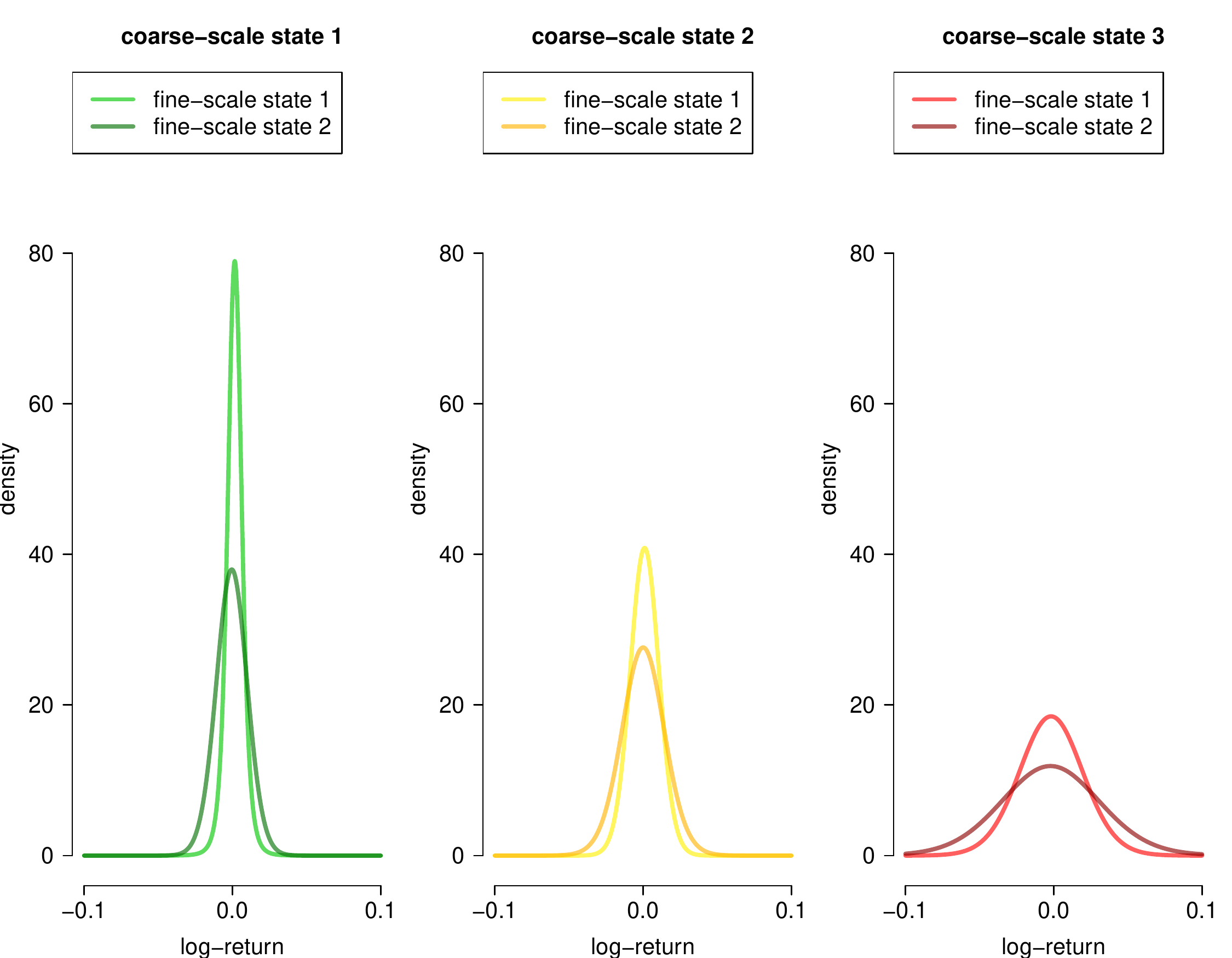}
		\label{fig:dax_statedepdistr}
	}
	\quad
	\subfigure[Pseudo-residuals of coarse-scale (top row) and fine-scale (bottom row) observations.]{
		\includegraphics[width=0.45\textwidth]{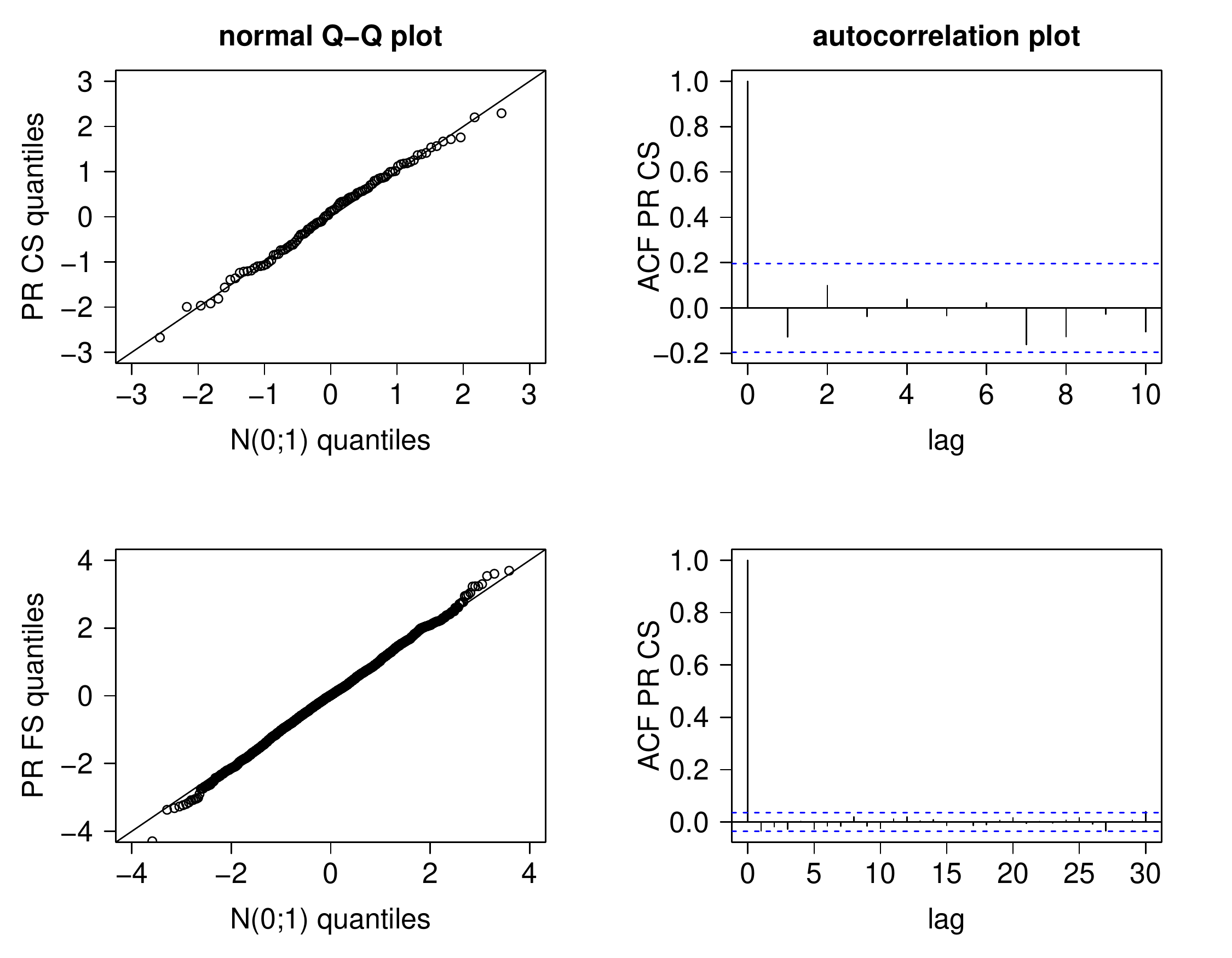}
		\label{fig:dax_pseudos}
	}
	\subfigure[Decoded time series of daily closing prices (top panel) and log-returns (bottom panel).]{
		\includegraphics[width=\textwidth]{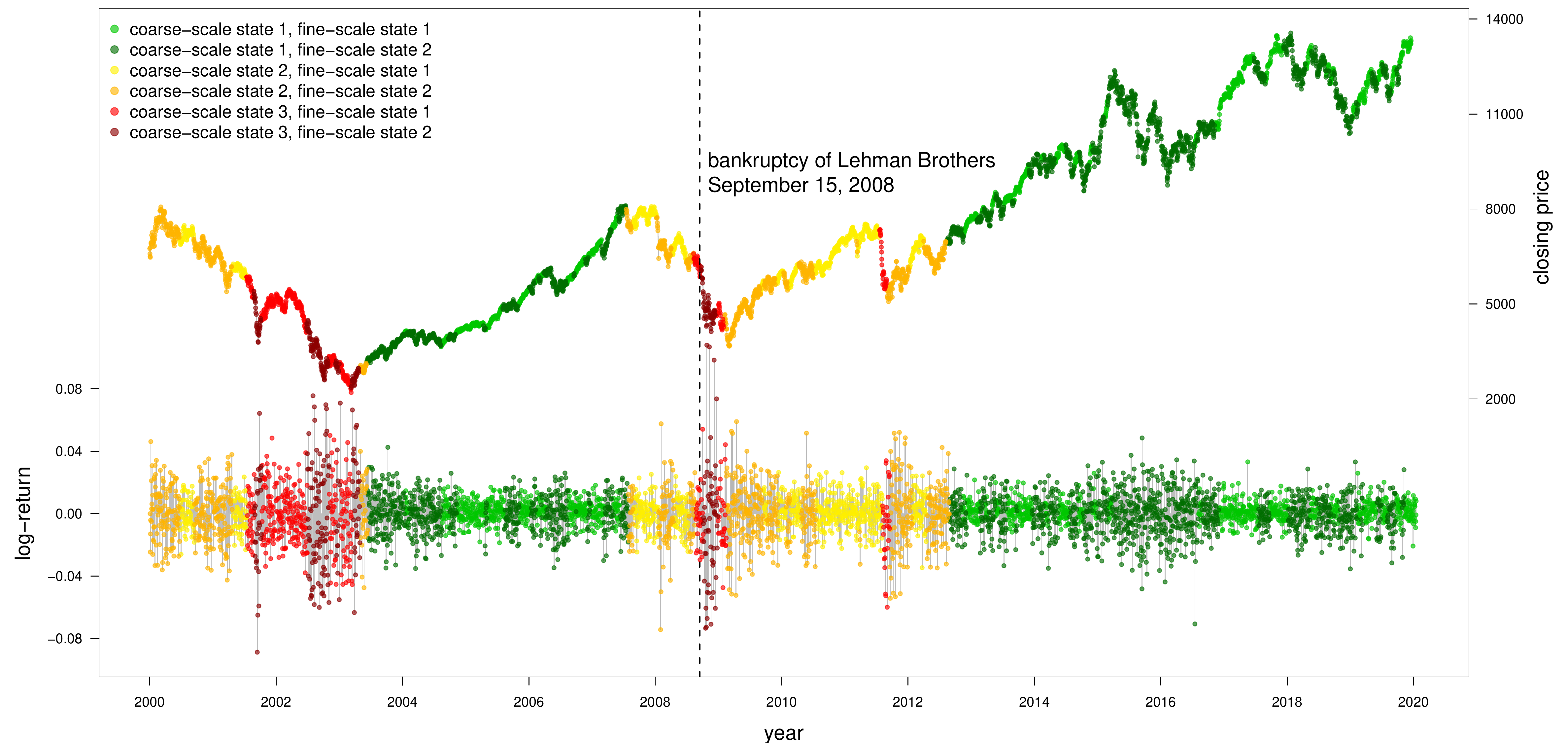}
		\label{fig:dax_decodedts}
	}
	\caption{Visualization of the HHMM results for the DAX.}  
	\label{fig:dax_results}
\end{figure}

The t.p.m.\ associated with the coarse-scale state process was estimated as
\begin{equation*}
\hat{\Gamma} = \begin{pmatrix}
0.936 & 0.053 & 0.011\\
0.073 & 0.828 & 0.099\\
0.000 & 0.319 & 0.681
\end{pmatrix},
\end{equation*}
which implies the stationary distribution $(0.458, 0.402, 0.140)$. The stationary state probabilities can be regarded as the long-term proportion of time that the coarse-scale state process spends in the different states. The t.p.m.s associated with the fine-scale state processes were estimated as
\begin{equation*}
\hat{\Gamma}_{1} = \begin{pmatrix}
0.924 & 0.076\\
0.070 & 0.930 
\end{pmatrix}, ~
\hat{\Gamma}_{2} = \begin{pmatrix}
0.923 & 0.077\\
0.077 & 0.923
\end{pmatrix}, ~
\hat{\Gamma}_{3} = \begin{pmatrix}
0.932 & 0.068\\
0.052 & 0.948
\end{pmatrix},
\end{equation*}
implying the stationary distributions $(0.479, 0.521)$, $(0.500,0.500)$, and $(0.433, 0.567)$, respectively. 

The estimated state-dependent scaled $t$-distributions of fine-scale log-returns are visualized in Figure \ref{fig:dax_statedepdistr}. Coarse-scale state 1 (bullish market) corresponds to the lowest marginal volatility ($7.5\cdot 10^{-3}$) and highest marginal expected return ($5.9\cdot 10^{-4}$), while coarse-scale state 3 (bearish market) corresponds to the highest marginal volatility ($28.1\cdot 10^{-3}$) and lowest marginal expected return ($-20.6\cdot 10^{-4}$). According to the stationary distribution under the fitted model, the market was in a bearish state in about 14.0 \% of the time, whereas the bullish market was active in about 45.8 \% of the time. Figure \ref{fig:dax_pseudos} shows that the pseudo-residuals can be considered as independently normal distributed, indicating a reasonable model fit. 

Figure \ref{fig:dax_decodedts} displays the decoded time series, with the decoding performed using the Viterbi algorithm as described in Section \ref{sec2:3}. In the autumn of 2008, the DAX is marked by the global financial crisis, which led to the bankruptcy of the US investment bank Lehman Brothers on September 15, 2008. It is noteworthy that the model detects fine-scale state 2 within the bearish market (which represents most lossy periods) and switches to calmer fine-scale states as soon as the log-returns become more moderate. In 2017, we observe a high proportion of light green periods, in which the DAX gained nearly 4{,}000 points within just a few months. In 2018, this skyrocketing was stopped. However, as the volatility remained low, the model retained coarse-scale state 1. 

\subsection{S\&P 500}
The American S\&P 500 index averages the stock performance of 500 companies listed on stock exchanges in the United States and is computed in a similar way as the DAX. We chose the same model setup as in Section \ref{sec3-1}, again considering 3 coarse-scale states and 2 fine-scale states. 

The t.p.m.\ associated with the coarse-scale state process was estimated as
\begin{equation*}
\hat{\Gamma} = \begin{pmatrix}
0.900 & 0.100 & 0.000\\
0.143 & 0.785 & 0.072\\
0.000 & 0.269 & 0.731
\end{pmatrix},
\end{equation*}
which implies the stationary distribution $(0.530, 0.371, 0.099)$. The t.p.m.s associated with the fine-scale state processes were estimated as
\begin{equation*}
\hat{\Gamma}_{1} = \begin{pmatrix}
0.955 & 0.045\\
0.023 & 0.977 
\end{pmatrix}, ~
\hat{\Gamma}_{2} = \begin{pmatrix}
0.969 & 0.031\\
0.031 & 0.969
\end{pmatrix}, ~
\hat{\Gamma}_{3} = \begin{pmatrix}
0.953 & 0.047\\
0.051 & 0.949
\end{pmatrix},
\end{equation*}
implying the stationary distributions $(0.338, 0.662)$, $(0.500, 0.500)$, and $(0.520, 0.480)$, respectively.

\begin{figure}[t]
	\centering
	\subfigure[Estimated state-dependent distributions of daily log-returns.]{
		\includegraphics[width=0.45\textwidth]{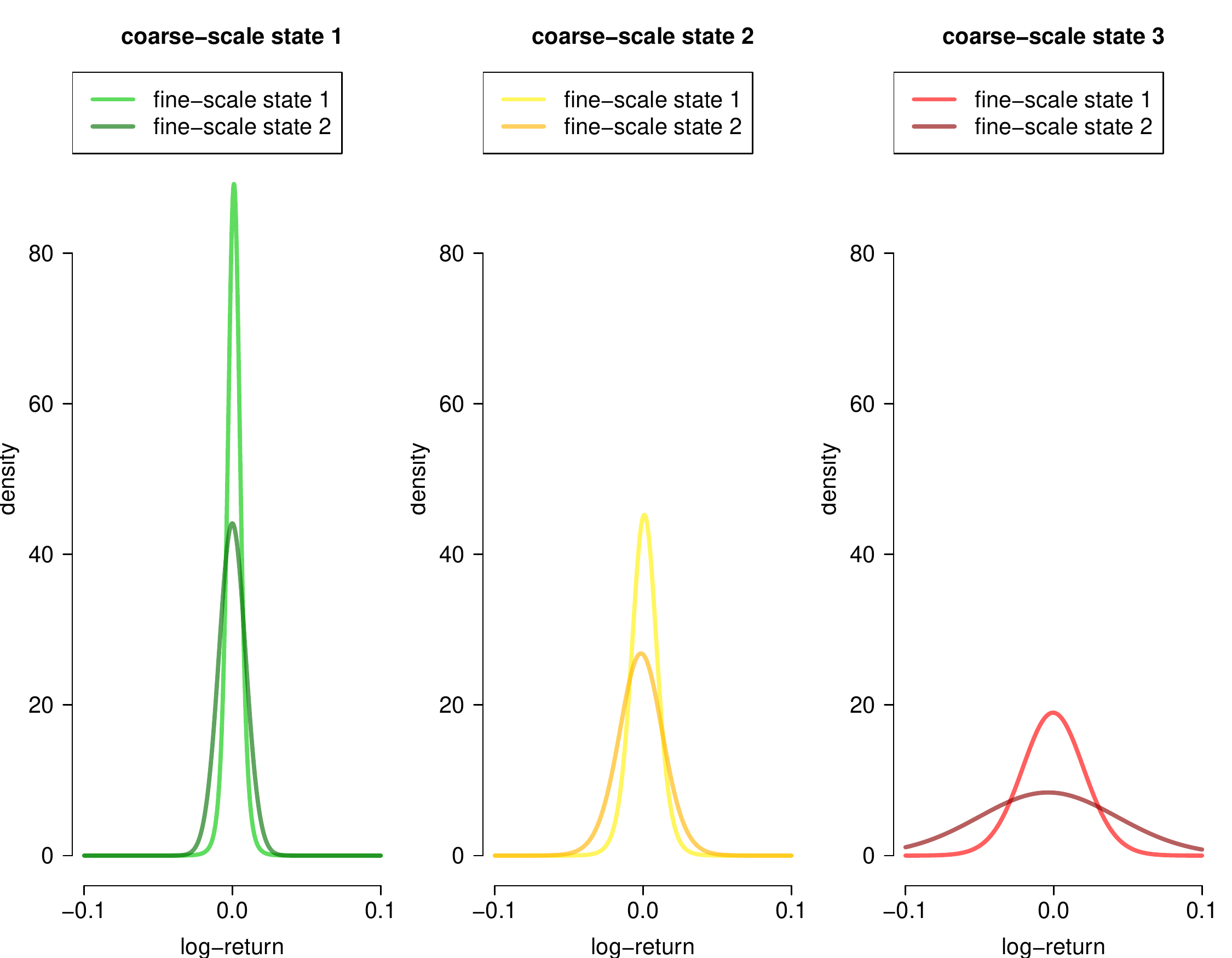}
		\label{fig:sandp_statedepdistr}
	}
	\quad
	\subfigure[Pseudo-residuals of coarse-scale (top row) and fine-scale (bottom row) observations.]{
		\includegraphics[width=0.45\textwidth]{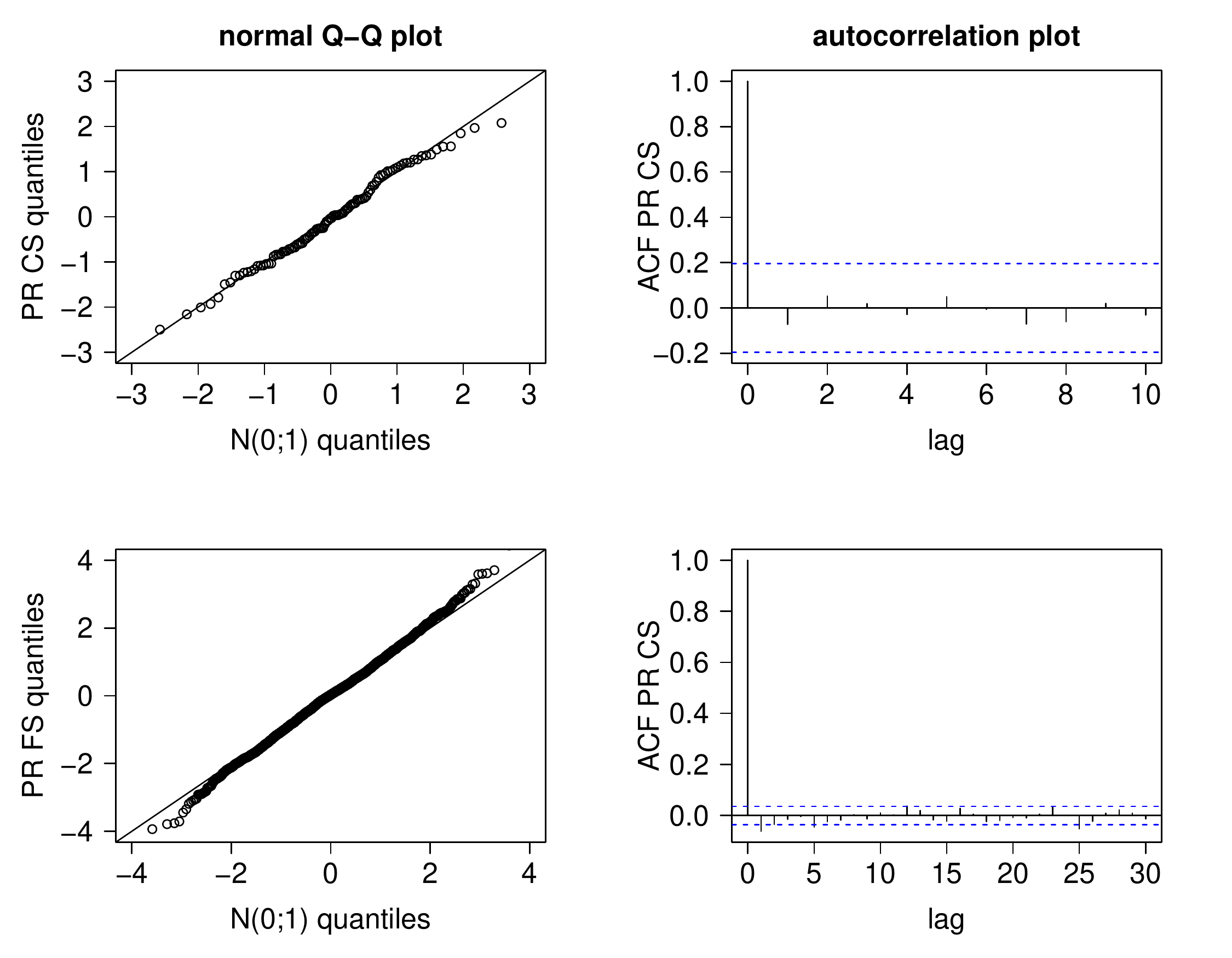}
		\label{fig:sandp_pseudos}
	}
	\subfigure[Decoded time series of daily closing prices (top panel) and log-returns (bottom panel).]{
		\includegraphics[width=\textwidth]{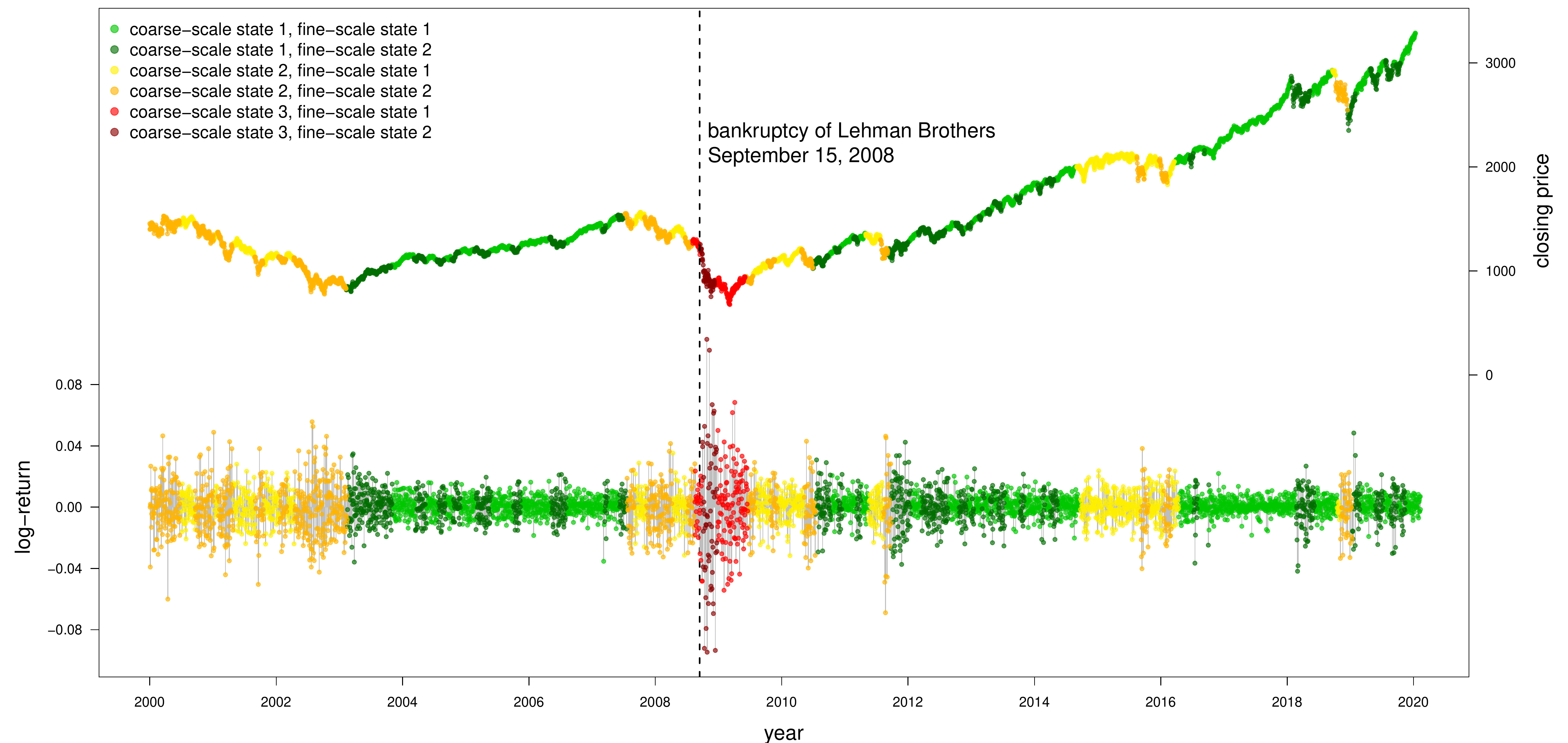}
		\label{fig:sandp_decodedts}
	}
	\caption{Visualization of the HHMM results for the S\&P 500.}  
	\label{fig:sandp_results}
\end{figure}

Further estimation results, as well as pseudo-residual checks, are visualized in Figure~\ref{fig:sandp_results}. These show very similar patterns to the DAX, with a slightly lower proportion (about 9.9 \%) of time spent in coarse-scale state 3 (bearish market). In fact, when looking at the decoded time series displayed in Figure \ref{fig:sandp_decodedts}, we observe that, in contrast to the DAX, the model retained coarse-scale state 2 throughout the early 2000s. 
In the autumn of 2008, however, the bankruptcy of Lehman Brothers was followed by a switch to coarse-scale state 3, which was then retained for several months, until a decrease in volatility led to a switch to coarse-scale state 2.

\section{Discussion} 
\label{sec4}

In this paper, we proposed HHMMs as a versatile extension of basic HMMs for detecting bearish and bullish markets in financial time series. 
By adding a hierarchical structure to the basic HMM, we paved the way for jointly inferring long-term trends and short-term dynamics from multi-scale time series data. In applications to two major stock indices, the DAX and the S\&P 500, we demonstrated that HHMMs can help to improve our understanding of how trends alternate, which in turn can help to make profitable investment decisions. In this last section, we discuss some limitations of the suggested approach and outline possible directions for future research.

First, the independence assumptions of the proposed model are to be questioned. Recall that we assumed conditional independence across observations and that the fine-scale HMMs are independent of the coarse-scale observations. While the former assumption is usually justified in the specific case of log-returns, the latter one can be regarded as more critical, as we used averages over the fine-scale data as coarse-scale observations. However, although inducing some dependence, we argue that averages of log-returns indicate changes in the long-term trend, which only depend on the current coarse-scale state and not on short-term fluctuations. 

Second, one can put into question whether financial market behavior can be classified into a finite number of states. For animal movement data, e.g., discrete states explaining resting, foraging, or traveling behavior seem natural. While the prospect of having similar proxies for financial data certainly is desirable, allowing for gradual changes may bring us closer to reality. In that regard, having discrete states on the coarse scale and a state-space model with an infinite number of states on the fine scale would be interesting to explore.  Since we should expect the interpretation of the states to become more complicated, such an extension can help us to overcome the deficit that discrete states are often prone to over-interpretation. 

On a final note, we would like to highlight that, in analogy to handwriting recognition, any desired number of hierarchies is theoretically feasible. Investigating whether such additional hierarchies, capturing e.g.\ medium-term trends or intra-day patterns, can help to draw a more complete picture of stock market behavior seems intriguing and is therefore to be regarded as a promising avenue for future research. 


\renewcommand\refname{References}
\makeatletter
\renewcommand\@biblabel[1]{}

\section*{Acknowledgements}

The authors wish to thank Roland Langrock for valuable discussions and helpful advice that considerably improved the paper.

\newpage
\section*{Appendix}
\label{sec5}

\begin{algorithm}[!htbp]
	\caption{Computing the log-likelihood of a fine-scale HMM.}
	\label{alg:llhmm}
	\begin{algorithmic}[1]
		\Procedure{$\log \mathcal{L}^{\text{HMM}}$}{$\theta^{*(i)}\mid (X^*_{t,t^*})_{t^*}$}
		\For{$k=1,\dots,N^*$}
		\State $\phi^{*(i)}_{k,1}=\log[\delta^{*(i)}_k]+\log[f^{*(i,k)}(X^*_{t,1})]$
		\EndFor
		\For{$t^*=2,\dots,T^*$}
		\State $c_{t^*-1} = \max\{ \phi^{*(i)}_{1,t^*-1},\dots,\phi^{*(i)}_{N^*,t^*-1} \}$
		\For{$k=1,\dots,N^*$}
		\State $\phi^{*(i)}_{k,t^*}=\log[f^{*(i,k)}(X^*_{t,t^*})]+\log\left[\sum_{j=1}^{N^*}\gamma^{*(i)}_{jk}\exp[\phi^{*(i)}_{j,t^*-1}-c_{t^*-1}]\right]+c_{t^*-1}$
		\EndFor
		\EndFor
		\State $c_{T^*} = \max\{ \phi^{*(i)}_{1,T^*},\dots,\phi^{*(i)}_{N^*,T^*} \}$
		\State \Return $\log \left[ \sum_{k=1}^{N^*}\exp[\phi^{*(i)}_{k,T^*}-c_{T^*}]\right]+c_{T^*}$
		\EndProcedure
	\end{algorithmic}
\end{algorithm}

\begin{algorithm}[!htbp]
	\caption{Computing the log-likelihood of a HHMM.}\label{alg:llhhmm}
	\begin{algorithmic}[1]
		\State $\theta = (\delta,\Gamma,(f^{(i)})_i)$ \Comment{initialize the coarse-scale parameters \, }
		\For{$i=1,\dots,N$} 
		\State $\theta^{*(i)}=(\delta^{*(i)},\Gamma^{*(i)},(f^{*(i,k)})_k)$ \Comment{initialize the fine-scale parameters\hspace*{3.75mm}\text{ }}
		\EndFor
		\Procedure{$\log \mathcal{L}^{\text{HHMM}}$}{$\theta,(\theta^{*(i)})_i\mid (X_t)_t,((X^*_{t,t^*})_{t^*})_t$}
		\For{$i=1,\dots,N$}
		\State $\phi_{i,1}=\log[\delta_i]+\log \mathcal{L}^{\text{HMM}}(\theta^{*(i)}\mid (X^*_{1,t^*})_{t^*})+\log[f^{(i)}(X_{1})]$ 
		\Comment{\Call{$\log\mathcal{L}^{\text{HMM}}$}{} is defined in Algorithm \ref{alg:llhmm} \hspace*{1.2mm}\text{ }}
		\EndFor
		\For{$t=2,\dots,T$}
		\State $c_{t-1} = \max\{ \phi_{1,t-1},\dots,\phi_{N,t-1} \}$
		\For{$i=1,\dots,N$}
		\State $\phi_{i,t}=\log \mathcal{L}^{\text{HMM}}(\theta^{*(i)}\mid (X^*_{t,t^*})_{t^*})+\log[f^{(i)}(X_t)]+\log\left[\sum_{j=1}^{N}\gamma_{ji}\exp[\phi_{j,t-1}-c_{t-1}]\right]+c_{t-1}$
		\EndFor
		\EndFor
		\State $c_{T} = \max\{ \phi_{1,T},\dots,\phi_{N,T} \}$
		\State \Return $\log\left[\sum_{i=1}^{N}\exp[\phi_{i,T}-c_{T}]\right]+c_{T}$
		\EndProcedure
	\end{algorithmic}
\end{algorithm} 

\begin{algorithm}[!htbp]
	\caption{Decoding the hidden states.}
	\label{alg:viterbi}
	\begin{algorithmic}[1]
		\State $\theta = (\delta,\Gamma,(f^{(i)})_i)$
		\Procedure{Viterbi}{$\theta, (X_t)_{t}$}
		\For{$i=1,\dots,N$}
		\State $\kappa_{i,1}=\log[\delta_i]+\log[f^{(i)}(X_1)]$
		\EndFor
		\For{$t=2,\dots,T$}
		\For{$i=1,\dots,N$}
		\State $\kappa_{i,t}=\max_j \left( \kappa_{j,t-1} + \log[\gamma_{ji}] \right)+\log[f^{(i)}(X_t)]$
		\EndFor
		\EndFor
		\State $\hat{S}_T = \argmax_i \kappa_{i,T}$ 
		\For{$t=T-1,\dots,1$}
		\State $\hat{S}_t = \argmax_i \kappa_{i,t}\gamma_{i\hat{S}_{t+1}}$
		\EndFor
		\State \Return $(\hat{S}_t)_t$
		\EndProcedure
	\end{algorithmic}
\end{algorithm}

\end{spacing}

\end{document}